\documentstyle[12pt]{article}
\normalbaselines

\begin{document}

\baselineskip=24pt

\bibliographystyle{unsrt}
\vbox {\vspace{6mm}}
\bigskip
\bigskip
\bigskip
\bigskip
\begin{center} {\bf SYMPLECTIC TOMOGRAPHY AS CLASSICAL\\
APPROACH TO QUANTUM SYSTEMS}
\end{center}

\begin{center}
{\it S. Mancini, V. I. Man'ko\footnote{On leave from Lebedev Physical
Institute, Moscow}
 and P.Tombesi }
\end{center}

\begin{center}
Dipartimento di Matematica e Fisica, Universit\`a di Camerino, I-62032
Camerino\\
and\\
Istituto Nazionale di Fisica della Materia
\end{center}

\bigskip
\bigskip
\bigskip

\begin{abstract}
By using a generalization of the optical tomography technique we describe the
dynamics of a quantum system in terms of equations for a purely classical
probability distribution
which contains complete information about the system.
\end{abstract}

\section{Introduction}\label{s1}

Due to the Heisemberg \cite{Heisemberg} and Schr\"odinger-Robertson
\cite{Schrodinger},
\cite{Robertson} uncertainty relation for the position and momentum in
quantum systems, does not
exist joint distribution function in the phase space. Nevertheless, a
permanent wish to understand
quantum mechanics in terms of classical probabilities leads to introduce
the so called
quasi-probability distributions, such as Wigner function \cite{Wigner},
Husimi Q-function
\cite{Husimi} and Glauber-Sudarshan P-function \cite{Glauber},
\cite{Sudarshan}. Later on a set of
$s$-ordered quasi-distributions \cite{CahillGlauber} unified these
quasi-probabilities into
one-parametric family. Even in the early days of quantum mechanics Madelung
\cite{Madelung}
observed that the modulus and the phase of wave function obey the
hydrodinamical classical
equations, and along this line the stochastic quantization scheme has been
suggested by Nelson
\cite{Nelson} to link the classical stochastic mechanics formalism with the
quantum mechanical
basic entities, such as wave function and propagator.   In
some sense, also the hidden variables \cite{Belinfante} was proposed to
relate the quantum
processes to the classical ones. Nevertheless, up to date there not exsist
a formalism which
consistently connects the "two worlds".

The discussed quasi-probabilities illuminated the similarities and
the differences between classical and quantum considerations, and they are
widely used as
instrument for calculations in quantum theory \cite{Scully},
\cite{Tatarsky}. However, they cannot
play the role of classical distributions, since for example, the Wigner
function and the
P-function may have negative values. Althought the Q-function is always
positive and
normalized, it does not describe measurable distributions of concrete
physical variables.

Using the formalism of Ref.
\cite{CahillGlauber}, Vogel and Risken \cite{VogelRisken} found an integral
relation between the
Wigner function and the marginal distribution for the measurable homodyne
output variable which
represents a rotated quadrature. This result gives the possibility of
measuring the quantum state,
and it is referred as optical homodyne tomography \cite{Raymer}.

In Ref. \cite{Man1} a symplectic tomography procedure was suggested to
obtain the Wigner
function by measuring the marginal distribution for a shifted and squeezed
quadrature, which
depends on extra parameters. In Ref. \cite{DarPaul} the formalism of Ref.
\cite{VogelRisken}
was formulated in invariant form, relating the homodyne output distribution
directly to the
density operator. In Ref. \cite{Man2} the symplectic tomography formalism
was also formulated
in this invariant form and it was extended to the multimode case. Thus, due
to the introduction
of quantum tomography procedure the real positive marginal distribution for
measurable
observables, such as rotated shifted and squeezed quadratures, turned out
to determine
completely the quantum states.

The aim of the present work is to formulate the standard quantum dynamics
in terms of the classical
marginal distribution of the measurable shifted and squeezed quadrature
components, used in the
symplectic tomography scheme. Thus we obtain an alternative formulation of
the quantum system
evolution in terms of evolution of real and positive  distribution function
for measurable
physical observables. We will show the connection of such "classical"
probability evolution
with the evolution of the above discussed quasi-prbability distributions.

Examples relative to states of harmonic oscillator and free motion will be
considered in
the frame of the given formulation of quantum mechanics.

\section{Density operator and distribution for shifted and squeezed quadrature}

In Ref. \cite{Man1} it was shown that, for the  generic linear combination
of quadratures, which
is a measurable observable $(\hbar=1)$
\begin{equation}\label{X}
\hat X=\mu \hat q+\nu\hat p+\delta,
\end{equation}
where $\hat q$ and $\hat p$ are the position and momentum respectively,
the marginal distribution
$w(X,\mu,\nu,\delta)$ (normalized with respect to the $X$ variable),
depending upon three extra
real parameters
$\mu,\nu,\delta$, is related to the state of the quantum system, expressed
in terms of its Wigner
function $W(q,p)$, as follows
\begin{equation}\label{w}
w(X,\mu,\nu,\delta)=\int e^{-ik(X-\mu q-\nu
p-\delta)}W(q,p)\frac{dkdqdp}{(2\pi)^2}.
\end{equation}
This formula can be inverted and the Wigner function of the state can be
expressed in terms of
the marginal distribution \cite{Man1}
\begin{equation}\label{W}
W(q,p)=(2\pi)^2s^2e^{isX}w_F(X,sq,sp,s),
\end{equation}
where $w_F(X,a,b,s)$ is the Fourier component of the marginal distribution
(\ref{w}) taken with
respect to the parameters $\mu,\nu,\delta$, i.e.
\begin{equation}\label{wF}
w_F(X,a,b,s)=\frac{1}{(2\pi)^3}\int w(X,\mu,\nu,\delta)e^{-i(\mu a+\nu
b+\delta s)}
d\mu d\nu d\delta.
\end{equation}
Hence, it was shown that the quantum state could be described by the
positive classical marginal
distribution for the squeezed, rotated and shifted quadrature. In the case
of only rotated
quadrature, $\mu=\cos\phi$, $\nu=\sin\phi$ and $\delta=0$, the usual
optical tomography formula of
Ref.
\cite{VogelRisken}, gives the same possibility  through the Radon transform
instead of the Fourier
transform. This is, in fact, a partial case of the symplectic
transformation of quadrature since
the rotation group is a subgroup of the symplectic group
$ISp(2,R)$ whose parameters are used to describe the transformation (\ref{X}).

In Ref. \cite{Man2} an invariant form connecting directly the marginal
distribution
$w(X,\mu,\nu,\delta)$ and the density operator was found
\begin{equation}\label{rho}
\hat\rho=\int d\mu d\nu d\delta\;w(X,\mu,\nu,\delta) \hat K_{\mu,\nu,\delta},
\end{equation}
where the kernel operator has the form
\begin{equation}\label{K}
\hat K_{\mu,\nu,\delta}=\frac{1}{2\pi}s^2e^{is(X-\delta)}
e^{-is^2\mu\nu/2}
e^{-is\nu \hat p}
e^{-is\mu \hat q}.
\end{equation}
The formulae (\ref{W}) and
(\ref{rho}) of symplectic tomography show that there exist an invertible
map between the quantum
states described by the set of nonnegative and normalized hermitian density
operators $\hat\rho$
and the set of positive, normalized marginal distributions ("classical"
ones) for the measurable
shifted and squeezed quadratures. So, the information contained in the
marginal distribution is
the same which is contained in the density operator; and due to this, one
could represent the
quantum dynamics in terms of evolution of the marginal probability.

\section{Quantum evolution as classical process}

We now derive the evolution equation for the marginal distribution function
$w$ using the invariant form of the connection between the marginal
distribution and the density operator given by the formula (\ref{rho}).
Then from the equation of
motion for the density operator
\begin{equation}\label{rhoH}
\partial_t\hat\rho=-i[\hat H,\hat\rho]
\end{equation}
we obtain the evolution equation for the marginal distribution in the form
\begin{equation}\label{wevo}
\int d\mu d\nu d\delta\; \left\{\dot w(X,\mu,\nu,\delta,t)\hat
K_{\mu,\nu,\delta}+
w(X,\mu,\nu,\delta,t)\hat I_{\mu,\nu,\delta}\right\}=0
\end{equation}
in which the known Hamiltonian determines the kernel $\hat
I_{\mu,\nu,\delta}$ through the
commutator
\begin{equation}\label{I}
\hat I_{\mu,\nu,\delta}=i[\hat H,\hat K_{\mu,\nu,\delta}].
\end{equation}
The obtained integral-operator equation for simple cases can be reduced to
the partial
differential equation.
To do this we represent the kernel operator $\hat I_{\mu,\nu,\delta}$ in
normal order form (i.e.
all the momentum operators on the left side and the position ones on the
right side)
containing the operator $\hat K_{\mu,\nu,\delta}$ as follow
\begin{equation}\label{Inormal}
:\hat I_{\mu,\nu,\delta}:={\cal R}(\hat p):\hat K_{\mu,\nu,\delta}:{\cal
P}(\hat q)
\end{equation}
where ${\cal R}(\hat p)$ and ${\cal P}(\hat q)$ are, finite or infinite
operator polynomials
(depending also on the parameters $\mu$ and $\nu$) determined by the
Hamiltonian.
Then calculating the matrix elements of the operator equation (\ref{wevo})
between the states
$\langle p|$ and $|q\rangle$ and using the completness property of the
Fourier exponents we
arrive at the following partial differential equation for the marginal
distribution function
\begin{equation}\label{weq}
\partial _t w+\Pi(\tilde p,\tilde q)w=0
\end{equation}
where the polynomial $\Pi(\tilde p,\tilde q)$ is the product of the
polynomials ${\cal R}(p)$
and ${\cal P}(q)$ represented in the form
\begin{equation}\label{Pi}
\Pi(p,q)={\cal R}(p){\cal P}(q)=\sum_n\sum_m p^nq^mc_{n,m}(\mu,\nu)
\end{equation}
in which the c-number variables $p$ and $q$ should be replaced by the operators
\begin{equation}\label{rep}
\tilde
p=\left(\frac{1}{\partial/\partial\delta}\frac{\partial}{\partial\nu}+i\frac
{\mu}{2}
\frac{\partial}{\partial\delta}\right),\qquad
\tilde
q=\left(\frac{1}{\partial/\partial\delta}\frac{\partial}{\partial\mu}+i\frac
{\nu}{2}
\frac{\partial}{\partial\delta}\right);
\end{equation}
where derivative in the denominator is understood as integral operator.
One should point out that the operators $\tilde p$ and $\tilde q$ in Eq.
(\ref{weq})
act on the product of coefficients  $c_{n,m}(\mu,\nu)$ and the marginal
distribution
corresponding to the order shown by Eqs. (\ref{weq}) and (\ref{Pi}). Let us
consider the  important
example of the particle motion in a potential with the Hamiltonian
\begin{equation}\label{HV}
\hat H=\frac{{\hat p}^2}{2}+V(\hat q);
\end{equation}
then the described procedure of calculating the normal order kernel
(\ref{Inormal}) gives the
following form of the quantum dynamics in terms of a Fokker-Planck-like
equation for the
marginal distribution
\begin{equation}\label{FPeq}
\dot
w-\mu\frac{\partial}{\partial\nu}w-
i\left[V\left(\frac{1}{\partial/\partial\delta}
\frac{\partial}{\partial\mu}+i\frac{\nu}{2}
\frac{\partial}{\partial\delta}\right)-
V\left(\frac{1}{\partial/\partial\delta}
\frac{\partial}{\partial\mu}-i\frac{\nu}{2}
\frac{\partial}{\partial\delta}\right)\right]w=0
\end{equation}
which in general case is an integro-differential equation.
For the free motion, $V=0$, this evolution  equation becomes the first
order partial differential
equation
\begin{equation}\label{freeeq}
\dot w-\mu\frac{\partial}{\partial\nu}w=0.
\end{equation}
For the harmonic oscillator, $V(\hat q)={\hat q}^2/2$, the quantum dynamic
equation has the form
\begin{equation}\label{hoeq}
\dot w-\mu\frac{\partial}{\partial\nu}w+\nu\frac{\partial}{\partial\mu}w=0.
\end{equation}
Thus given a Hamiltonian  of the form (\ref{HV}) we can study the quantum
evolution of
the system writing down a Fokker-Planck-like equation for the marginal
distribution. Solving this
one for a given initial positive and normalized marginal distribution we
can obtain the
quantum density operator $\hat\rho(t)$ according to Eq. (\ref{rho}).
Conceptually it means that
we can discuss the system quantum evolution considering classical real
positive and normalized
distributions for the measurable variable $X$ which is shifted and squeezed
quadrature.
The distribution function which depends on extra parameters obeys a classical
equation which preserves the normalization condition of the
distribution. In this sense we always can reduce the quantum behaviour of
the system to the
classical behaviour of the marginal distribution of the shifted and
squeezed quadrature. Of
course, this statement respects the uncertainty relation because the
measurable marginal
distribution is the distribution for one observable. That is the essential
difference (dispite of
some similarity) of the introduced marginal distribution from the discussed
quasi-distributions,
including  the real positive Q-function, which depend on the two variables
of the phase space and
are normalized with respect to these variables. We would point out that we
do not derive
quantum mechanics from classical stochastic mechanics, i.e. we do not
quantize any classical
stochastic process, our result is to present the quantum dynamics equations
as classical
ones, and in doing this we need not only classical Hamiltonian but also its
quantum
counterpart.

\section{Examples}

Below we consider simple examples of the marginal distribution evolution
for states of free
motion and harmonic oscillator.
First of all we take into account the free motion for which the Eq.
(\ref{freeeq}) has a gaussian
solution of the form
\begin{equation}\label{freesolution}
w(X,\mu,\nu,\delta,t)=
\frac{1}{\sqrt{2\pi\sigma_X(t)}}\exp\left\{-\frac{(X-\delta)^2}
{2\sigma_X(t)}\right\}
\end{equation}
where the dispersion of the observable $X$ depends on time and parameters
as follow
\begin{equation}\label{sigma}
\sigma_X(t)=\frac{1}{2}[\mu^2+\nu^2(1+t^2)+2\mu\nu t].
\end{equation}
The initial condition corresponds to the marginal distribution of the
ground state of an
artificial harmonic oscillator calculated from the respective Wigner
function \cite{Man1}.

If we consider the first excited state of the harmonic
oscillator, we know the Wigner function \cite{Gardiner}
\begin{equation}\label{W1}
W_1(q,p)=-2(1-2q^2-2p^2)\exp[-q^2-p^2].
\end{equation}
It results time independent due to the stationarity of the state, but for
small $q$ and $p$ it
becomes negative while the solution of Eq. (\ref{hoeq})
\begin{equation}\label{w1}
w_1(X,\mu,\nu,\delta,t)=
\frac{2}{\sqrt{\pi}}[\mu^2+\nu^2]^{-\frac{3}{2}}(X-\delta)^2
\exp\left\{-\frac{(X-\delta)^2}{\mu^2+\nu^2}\right\}
\end{equation}
is itself time independent, but everywhere positive.

Indeed, a time evolution is present explicitly in
the coherent state, whose Wigner function is given by
\begin{equation}\label{Wc}
W_c(q,p)=2\exp\{-q^2-q_0^2-p^2-p_0^2+2(qq_0+pp_0)\cos t-(pq_0-qp_0)\sin t\}
\end{equation}
where $q_0$ and $p_0$ are the initial values of position and momentum.
For the same state the marginal distribution shows a more complicate evolution
\begin{eqnarray}\label{wc}
&w_c&(X,\mu,\nu,\delta,t)=\frac{1}{\sqrt{\pi}}[\mu^2+\nu^2]^{-\frac{1}{2}}\\
&\times&\exp\left\{-q_0-p_0-\frac{(X-\delta)^2}{\nu^2}
+2\frac{(X-\delta)}{\nu}(p_0\cos t-q_0\sin
t)\right\}\nonumber\\
&\times&\exp\left\{\frac{1}{\mu^2+\nu^2}\left[\frac{\mu}{\nu}(X-\delta)+
q_0(\mu\sin t+\nu\cos t)
+p_0(\nu\sin t-\mu\cos t)\right]^2\right\}.\nonumber
\end{eqnarray}
It is also interesting to consider the comparison between Wigner function
and marginal
probability for non-classical states of the harmonic oscillator, such as
female cat state
defined as \cite{Manko}
\begin{equation}\label{cat}
|\alpha_{-}\rangle=N_{-}(|\alpha\rangle-|-\alpha\rangle),\quad\alpha=2^{-1/2
}(q_0+ip_0)
\end{equation}
with
\begin{equation}\label{N_}
N_{-}=\left\{\frac{\exp[(q_0^2+p_0^2)/2]}{4\sinh[(q_0^2+p_0^2)/2]}\right\}^{
\frac{1}{2}}
\end{equation}
and for which the Wigner function assumes the following form
\begin{eqnarray}\label{Wcat}
W_{-}(q,p)=2N_{-}^2e^{-q^2-p^2}&\{&e^{-q_0^2-p_0^2}\cosh[2(qq_0+pp_0)\cos
t+2(qp_0-pq_0)\sin t]
\nonumber\\
&-&\cos[2(qp_0-pq_0)\cos t-2(qq_0+pp_0)\sin t]\}.
\end{eqnarray}
The corresponding marginal distribution is
\begin{eqnarray}\label{wcat}
w_{-}(X,\mu,\nu,\delta,t)&=&N_{-}^2[w_A(X,\mu,\nu,\delta,t)-
w_B(X,\mu,\nu,\delta,t)
\nonumber\\
&-&w_B^*(X,\mu,\nu,\delta,t)-w_A(-X,\mu,\nu,-\delta,t)]
\end{eqnarray}
with
\begin{eqnarray}
&w_A&(X,\mu,\nu,\delta,t)=
\frac{1}{\sqrt{\pi}}[\mu^2+\nu^2]^{-\frac{1}{2}}\\
&\times&\exp\left\{-q_0-p_0-
\frac{(X-\delta)^2}{\nu^2}+
2\frac{(X-\delta)}{\nu}(p_0\cos t-q_0\sin t)\right\}\nonumber\\
&\times&\exp\left\{\frac{1}{\mu^2+\nu^2}\left[\frac{\mu}{\nu}(X-\delta)+
q_0(\mu\sin t+\nu\cos t)
+p_0(\nu\sin t-\mu\cos t)\right]^2\right\}\nonumber
\end{eqnarray}
and
\begin{eqnarray}
&w_B&(X,\mu,\nu,\delta,t)=\frac{1}{\sqrt{\pi}}[\mu^2+\nu^2]^{-\frac{1}{2}}\\
&\times&\exp\left\{-q_0-p_0-\frac{(X-\delta)^2}{\nu^2}
-2i\frac{(X-\delta)}{\nu}(q_0\cos t+p_0\sin
t)\right\}\nonumber\\
&\times&\exp\left\{\frac{-1}{\mu^2+\nu^2}\left[-i\frac{\mu}{\nu}(X-\delta)+
q_0(\mu\cos t-\nu\sin t)
+p_0(\mu\sin t+\nu\cos t)\right]^2\right\}.\nonumber
\end{eqnarray}
The presented examples show that for the evolution of the state of a
quantum system, one
could always associate the evolution of the probability density for the
random classical variable
$X$ which obeys "classical" Fokker-Planck-like equation, and this
probability density contains the
same information (about quantum system) which is contained in any
quasi-distribution function. But
the probability density has the advantage to behave completly as the usual
classical one. The
physical meaning of the "classical" random variable $X$ is transparent, it
is considered as the
position in an ensamble of shifted, rotated and scaled rest frames in the
classical phase
space of the system under study. We could remark that for non normalized
quantum states, like the
states with fixed momentum (De Broglie wave) or with fixed position,
the introduced map in Eq. (\ref{rho}) may be  preserved. In this context
the plane
wave states of free motion have the marginal distribution corresponding to
the classical white
noise.

\section{Conclusions}

We have shown that it is possible to bring the quantum dynamics back to
classical description in
terms of a probability distribution containing (over)complete information.
The time evolution of
a measurable probability for the discussed observables could be useful both
for the prediction of
the experimental outcomes  at a given time and, as mentioned above, to
achieve the quantum state
of the system at any time.
Furthermore the symplectic transformation of Eq. (\ref{X}) could be
represented as a composition
of shift, rotation and squeezing.
So, the measurement of a shifted
variable means the measure of the coordinate in a frame in which the zero
is shifted. This could be
implemented for example by measuring the oscillator coordinate using an
infinite ensamble of
frames which are shifted with respect to the initial one (related method
was discussed also in
Ref. \cite{Royer}). Furthermore if one considers the variable $\hat q$ as
the photon quadrature,
which corresponds to the amplitude of the electric vector vibrations, a
rotation means a homodyne
measurement, while the squeezing means measurement after amplification or
attenuation. So, we
would emphasize that our procedure allows to transform the problem of
quantum measurements (at
least for some observables) into a problem of classical  measurements with
an ensamble of shifted,
rotated and scaled reference frames in the (classical) phase space.

We also want to remark that in some situations the measurements of
instataneous values of the
marginal distribution for different values of the parameters is replaced by
measuring the
distribution for these parameters which evolve in time. Such measurements
may be consistent with
the system evolution if the parameters time variation is much faster than
the natural
evolution of the system itself. In this case the state of the system does
not change during
the measurement process and one obtains the instant value of the marginal
distribution and of
the Wigner function.

Finally we belive that our "classical" approach could be a powerful tool to
investigate complex quantum system as for example chaotic systems in which
the quantum chaos could
be considered in a frame of equations for a real and positive distribution
function.

\section*{\bf Acknowledgments}

This work has been partially supported by European Community under the
Human Capital and Mobility (HCM) programme. V. I. M. would also like to
acknowledge the University
of Camerino for kind hospitality.

\end{document}